\documentstyle[pre,aps,epsf,floats]{revtex}

\begin{document} 

\flushbottom

\draft
\twocolumn[\hsize\textwidth\columnwidth\hsize\csname @twocolumnfalse\endcsname

\title{Critical behavior of a traffic flow model}

\author{L.~Roters\cite{LarsEmail}, S. L\"ubeck\cite{SvenEmail}, and
K.~D.~Usadel\cite{UsadelEmail} }
\address{
Theoretische Physik, 
Gerhard-Mercator-Universit\"at Duisburg, 
Lotharstra{\ss}e 1, 47048 Duisburg, Germany}

\date{Received 08 July 1998}

\maketitle

\begin{abstract}
The Nagel-Schreckenberg traffic flow model
shows a transition from a free flow regime
to a jammed regime for increasing car density.
The measurement of the dynamical structure factor
offers the chance to observe the evolution of jams 
without the necessity to define a car to be jammed
or not.
Above the jamming transition the dynamical structure
factor exhibits for a given $k$-value two maxima 
corresponding to the separation of the system into the 
free flow phase and jammed phase.
We obtain from a finite-size scaling analysis of 
the smallest jam mode that approaching the
transition long range correlations of the jams
occur.
\end{abstract}

\pacs{89.40.+k,05.40.+j,05.60.+w}

\narrowtext

]  

\setcounter{page}{1}
\markright{\rm
Phys. Rev.~E {\bf 59}, ? (1999), accepted for publication. }
\thispagestyle{myheadings}
\pagestyle{myheadings}

\section{Introduction}
\label{sec:intro}

Real traffic displays with increasing car density a transition from
free flow traffic to congested traffic. 
This break down of free flow traffic is accompanied by the occurrence of
traffic jams which are 
not attributable to any external cause but only to
distance fluctuations between following vehicles within very dense and
unstable traffic (see for instance~\cite{LEUTZBACH_1}). 
These traffic jams find expression in shock waves, i.e., in backward moving
density fluctuations. 
This property of jams was already found during the 60's in traffic
observations~\cite{TREITERER_1}. 
Beyond this early experimental investigations recently performed 
measurements of real highway traffic 
leads to the conjecture that jams are 
characterized by several independent parameters, e.g., the 
jam velocity, mean flux, etc.~\cite{KERNER_1}.

Since the seminal work of Lighthill and Whitham 
in the middle of the 50's~\cite{LIGHT_1} many attempts have been
made to construct more and more sophisticated models which incorporate
various phenomena occurring in real 
traffic (for an overview see~\cite{WORKSHOP_95,WORKSHOP_97}).
Few years ago Nagel and \mbox{Schreckenberg}~\cite{NASCH_1} introduced a 
cellular automata model, which simulates single-lane one-way traffic,
and which is able to reproduce the main features of traffic flow, 
backward moving shock waves and the so-called fundamental
diagram~(see \cite{HALL_1} and references therein).  
Investigations of the Nagel-Schreckenberg
traffic flow model show that crossing a
critical point a transition takes place from a 
homogeneous regime (free flow phase) to an inhomogeneous 
regime which is characterized by a coexistence of
the free flow phase and the jammed phase~\cite{LUEB_6,CHOWD_1}.
Thereby, the free flow phase is characterized by a low local
density and the jammed phase by a high
local density, respectively.
Due to the particle conservation of the model
the transition is realized by the system separating
into a low density region 
and a high density region~\cite{LUEB_6}.
Despite several attempts it is still an open question wether the
transition can be described as a critical phenomenon~(see for
instance~\cite{LUEB_6,VILAR_1,CSANYI_1,SASVARI_1,EISEN_2}). 

An attempt to investigate the occurrence of jam
was made by Vilar and Souza~\cite{VILAR_1} who introduced
an order parameter which equals the number of standing cars.  
Without noise this quantity vanishes continuously below the
transition. 
But including noise the amount of standing cars
is finite for any nonzero density,
i.e.,~the average number of standing cars does not vanish
below the critical density.
The crucial point is that below the transition a particle
may stop due to the noisy slowing down rule 
of the dynamics, 
but this behavior does not coincide with
backward moving density fluctuations and
algebraic decaying correlation functions which would indicate 
a critical behavior of the system.
At present, a convincing definition of an order
parameter, which tends below the critical density to zero,
is not known.
Again, it is even not clear wether the Nagel-Schreckenberg traffic
flow model displays criticality at all.

We use the dynamical structure factor both to
make predictions about the properties of the different phases 
observed in the  Nagel-Schreckenberg model and to investigate the
question wether the occurrence of jam can be described as a phase
transition or not.  
The dynamical structure factor which is closely related to the
correlation function is an appropriate tool to do this
because it naturally distinguishes between the two phases
characterized by positive and negative velocities
(free traffic flow and backward moving shock waves).  
Thus the advantage of our method of analyses is that
the properties of both phases can be examined without the
necessity to define a car to be jammed or not. 
The paper is organized as follows: 
In Sec.~II we briefly describe our analysis of the
dynamical structure factor and recall the main
results which were published recently~\cite{LUEB_7A}. 
In Sec. III we extend this method of analysis to
the so-called velocity-particle space and
address the question whether the transition from
the free flow regime to the phase coexisting
regime (where the system separates into the
jammed and free flow phase) can be considered
as a phase transition.
Basing on the dynamical structure factor, our results suggest
that a continuous phase transition takes place.

\section{Model and Simulations}
\label{sec:M_und_S}

The Nagel-Schreckenberg traffic flow model~\cite{NASCH_1}
is based on a one-dimensional cellular automaton of
linear size $L$ and $N$ particles.
Integer values describing
the position $r_n \in\{ 1,2,...,L \} $,
the velocity $v_n \in \{0,1,...,v_{\rm max} \} $
and the gap $g_n$ to the forward neighbor are associated with each 
particle. 
For each particle, the following 
update steps representing the acceleration, the slowing down, the noise,
and the motion of the particles are done in parallel:
(1) if \mbox{$v_n < g_n$} the velocity is increased with
respect to the maximal velocity,
$v_n \to \mbox{Min}\{v_n+1, v_{\rm max}\}$,
(2) to avoid crashes the velocity is decreased by $v_n \to g_n$ 
if $v_n > g_n$,
(3) if $v_n>0$ the velocity is decreased by $v_n \to v_n-1$ 
with probability $P\/$ in order to allow fluctuations, 
and finally (4) the motion of the cars is given by $r_n \to r_n+v_n$.
Thus the behavior of the model is determined by three parameters, the
maximal velocity $v_{\rm max}$, the noise parameter $P$ and the global
density of cars $\rho=N/L$, where $N$ denotes the total number of cars
and $L$ the system size, which is chosen to be $L=32768$ throughout
the whole paper.

Our analyses is based on the dynamical structure factor whose
definition is as follows: 
Consider the occupation function 
\begin{equation}
  \eta_{r,t} \; = \; \left \{ 
    \begin{array}{l}
      1 \hspace{0.5cm} \mbox{if cell $r$ is occupied at time $t$}\\
      0 \hspace{0.5cm} \mbox{otherwise} \, .
    \end{array} \right.
\end{equation}
The evolution of $\eta_{r,t}$ leads directly to the space-time diagram
where the propagation of the particles can be visualized
(see for instance Fig.~2 in~\cite{ITO}). 
The dynamical structure factor $S(k, \omega)$ is then given by
\begin{equation}
  S(k,\omega) \; = \; \frac{1}{l\,T}  \;\left \langle 
    \left | \;{\displaystyle\sum_{r,t}} \, \eta_{r,t} \; 
      e^{i (k r -\omega t) } \right |^2
  \right\rangle \, ,
  \label{eq:def_steady_state}
\end{equation}
where the Fourier transform is taken over a finite rectangle of the
space-time diagram of size $l\times T$, i.e.,~$r$ and $t$ are 
integers ranging from $1$ to $l$ and $1$ to $T$, respectively. 
Then $k$ and $\omega$ are also discrete values 
$k=2 \pi m_k/l$ and $\omega= 2 \pi m_{\omega}/T$ 
with $m_k\in\{0,1,2,...,l-1\}$ and 
$m_{\omega}\in\{0,1,2,...,T-1\}$, respectively.
The dynamical structure factor $S(k,\omega)$ is related
to the Fourier transform of the real space density-density
correlation function~$C(r,t)$ and compared to the
analysis of the steady state structure factor~\cite{LUEB_6}
and the related steady state correlation function~\cite{EISEN_2}
it contains both the spatial and the temporal evolution 
of the system.
Figure~\ref{fig:dyn_struc_factor} shows the dynamical structure
factor both below and above the transition. 
Below the transition $S(k, \omega)$ exhibits one mode formed by the
ridges. 
This mode is characterized by a positive slope 
$v={\partial \omega}/{\partial k}$ corresponding 
to the positive velocity $v_{\rm f}$ of the particles in the
free flow phase. 
Increasing the global density, a second mode
appears at the transition to the coexistence regime. 
This second mode exhibits a negative slope~($v_{\rm j}<0$)
indicating that it corresponds to the backward moving 
density fluctuations in the jammed phase.

Due to the sign of their characteristic velocities $v_{\rm f}$ and 
$v_{\rm j}$, both phases can clearly be distinguished. 
Recently performed investigations turned out that 
the characteristic velocity of the free flow phase 
equals the velocity ($v_{\rm f}=v_{\rm max}-P$) of free flowing cars 
in the low density limit ($\rho \to 0$), i.e., cars of the 
free flow phase behave as independent particles for all
densities~\cite{LUEB_7A}. 
The velocity of the jams neither depends on the global density nor on
the maximum velocity, i.e., it is a function of the noise parameter
only~\cite{LUEB_7A}. 
This result coincides with those
of recently published investigations~\cite{NEUBERT_LEE},
which were obtained from a variation analysis of a
multi-point-autocorrelation function.   
The jam velocity is, beside the maximum flow, 
the fluxes inside and outside of the jam and the average vehicle
speeds (see for instance~\cite{KERNER_1,KERNER_2} and
references therein) a characteristic parameter
of real traffic and its knowledge is therefore necessary to 
calibrate any traffic model to the conditions observed in real
traffic.

\section{The jammed phase}
\label{sec:jammed_phase}

\begin{figure}[b]
 \epsfxsize=8.6cm
 \epsffile{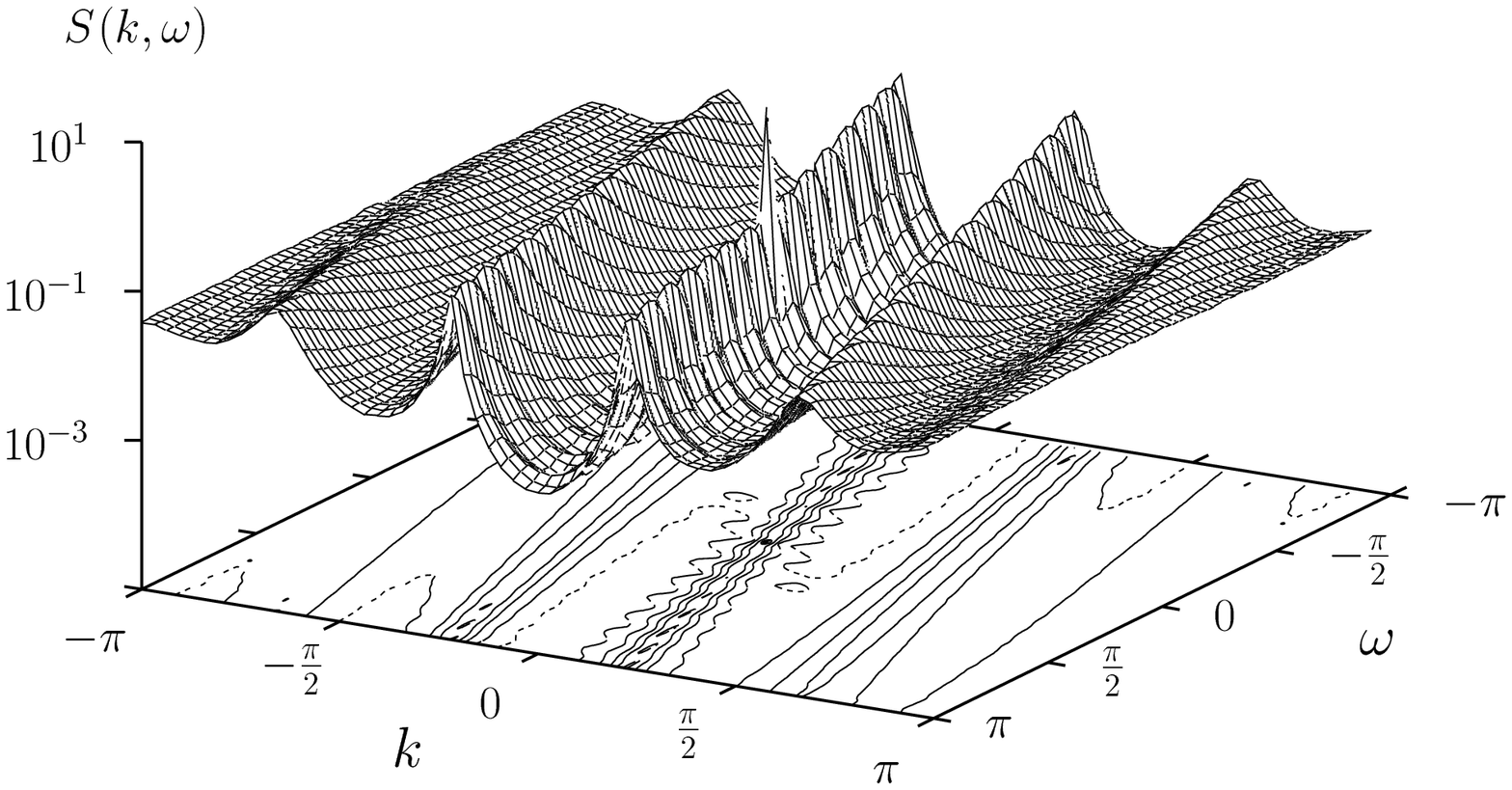}
 \epsfxsize=8.6cm
 \epsffile{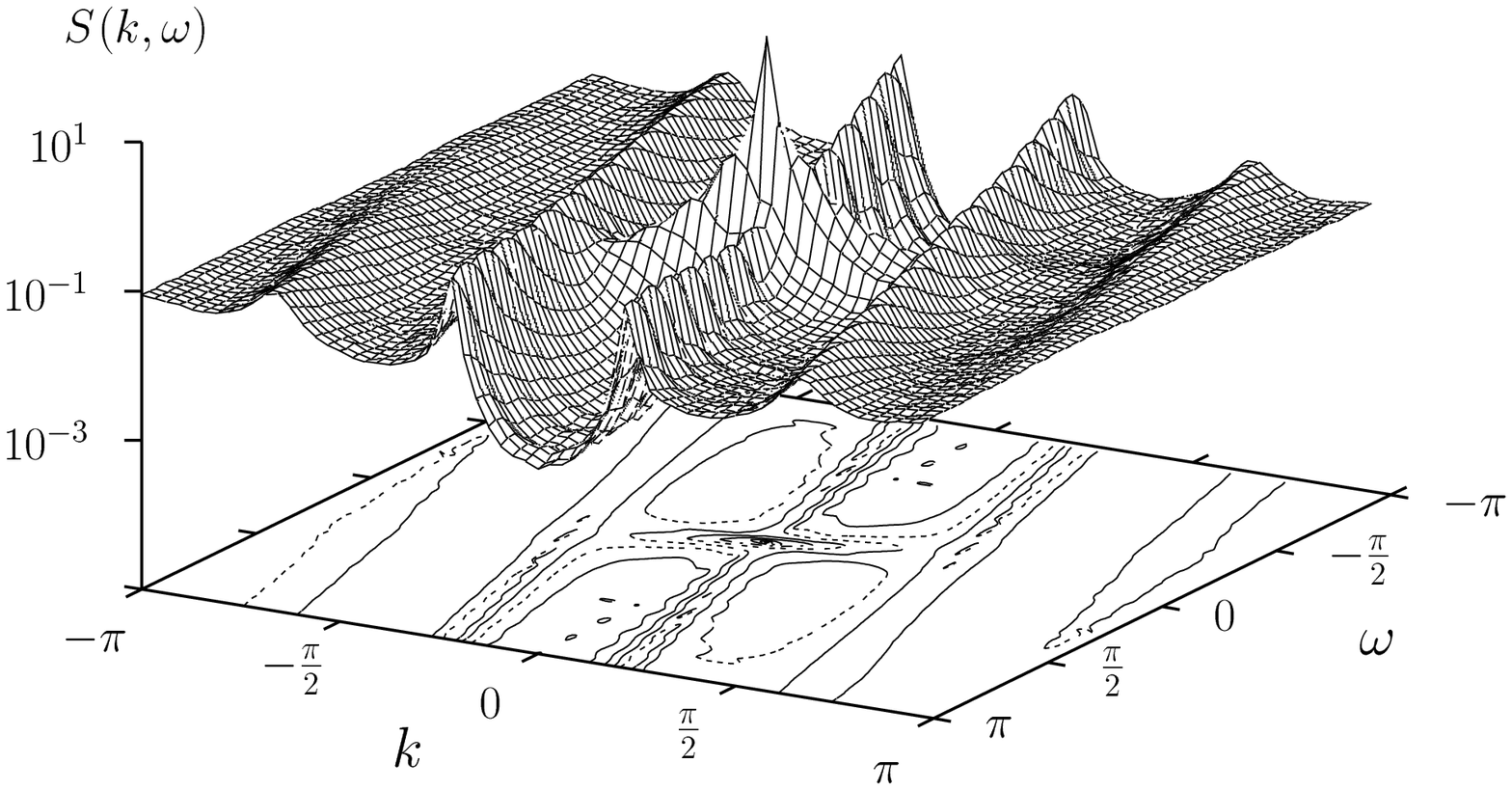}
  \caption{
    The dynamical structure factor $S(k,\omega)$ below (upper figure)
    and above (lower figure) the jamming transition.
    The ridges with maximal $S(k,\omega)$ 
    indicate the various modes.
    }
 \label{fig:dyn_struc_factor} 
\end{figure}

\begin{figure}[t]
 \epsfxsize=8.6cm
 \epsffile{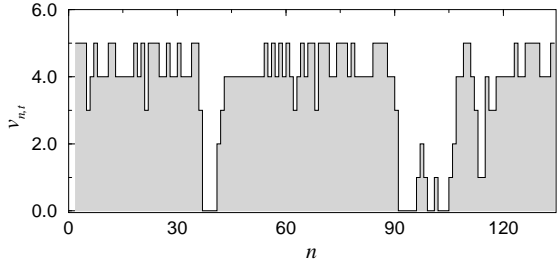}
  \caption{Snapshot of the velocity $v_{n,t}$ as a function
    of car number $n$ at a certain time $t$ and $P=0.5$. 
    Jams are characterized by 
    several particles with low velocity. 
    The free flow particles fluctuate independently between
    $v_{n,t}=v_{\rm max}-1$ and $v_{n,t}=v_{\rm max}$.}
 \label{fig:v_n_space} 
\end{figure}

In the last section we briefly described that the 
analysis of the dynamical structure factor allows
to determine the characteristic velocities of the jammed and the
free flow phase.
In this section we are interested in the correlations 
occurring within the jammed phase. 
Therefore it is necessary to analyse the jammed mode. 
Unfortunately, the superposition of the jammed and the free flow mode
makes it difficult to consider the pure jammed mode (see
Fig.~\ref{fig:dyn_struc_factor}).
Thus a refinement of the analysis of the dynamical
structure that allows to separate both modes
is needed.
This can be achieved by
changing from the occupation function $\eta_{r,t}$, defined in real
space, to the  velocity-particle space where the evolution of each
particle velocity $v_{n,t}$ is considered. 
A snapshot of the velocity-particle space for a given time is 
shown in Fig.~\ref{fig:v_n_space}.
In the free flow phase, where
the cars can be considered as independent
particles~\cite{LUEB_6}, the velocities fluctuate
according to the noisy slowing down rule 
between the two
values $v_{\rm max}$ and $v_{\rm max}-1$,
respectively. 
Extended regions with small or even zero 
velocities corresponds to traffic jams.
Comparable to the space-time diagram
these regions move backward in time.
The Fourier transform of the velocity-particle
diagram in two dimensions leads to the 
dynamical structure factor $S_v(k,\omega)$
which is given by 
\begin{equation}
  S_v(k,\omega) \; = \; \frac{1}{N\,T} \, \left \langle 
    \left | \sum_{n,t} \, v_{n,t} \, e^{i (k n -
    \omega t) } \right |^2 
\right\rangle ,
\label{eq:struc_fact_v}
\end{equation}
with $k= 2 \pi m_k/N$,
$\omega=2 \pi m_{\omega}/T$ and
where the $m$'s being integers.

\begin{figure}[b]
 \epsfxsize=8.6cm
 \epsffile{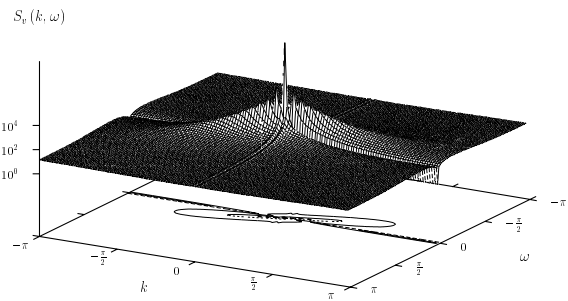}
  \caption{The dynamical structure factor $S_v(k, \omega)$
    of the velocity-particle space
    above the transition for $P=0.5$.
    Notice that the free flow phase contributes
    to $S_v(k, \omega)$ only white noise.
    \label{fig:dyn_struc_factor_v}} 
\end{figure}

Compared to the dynamical structure factor
of the ordinary space-time diagram 
the dynamical structure factor of the
velocity-particle space
has the advantage that the free flow phase
contributes only white 
noise ($S_v(k,\omega)|_{\rm free \; flow} = {\rm const}$), 
i.e.,~the analysis of the
occurring traffic jams is made easier.
Figure~\ref{fig:dyn_struc_factor_v} shows
the structure factor $S_v(k, \omega)$ above
the transition.
The structure factor displays one ridge
with a negative slope, corresponding to the backward
moving jams.
The notch parallel to the $k$-axis through $\omega=0$ 
is caused by finite-size effects and disappears with
increasing system size.
The peak in the center of the diagram [$S_v(k=0,\omega =0)$]
describes the velocity fluctuations of the whole system.
It is an average over both coexisting phases and therefore it leads to
uncertain results on the occurrence of jams.

\begin{figure}[t]
 \epsfxsize=8.6cm
 \epsffile{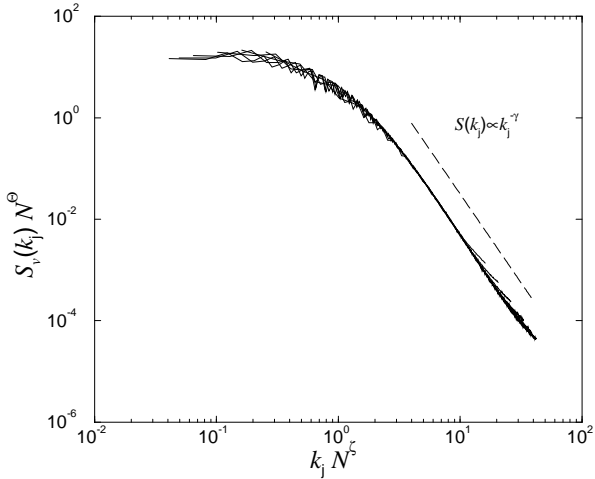}
  \caption{The scaling plot of $S_v(k, \omega)$
    along the jam mode for various values of
    $N$ (with $N\in\{128, 256, 512, 1024, 2048\}$ and $N=T$) 
    and a fixed value of the global density with $\rho>\rho_c$.
    For $\theta = -1.22$ and $\zeta = 0.34$ we got a 
    good data collapse indicating that the correlation
    of the jams decay algebraically
    (dashed line with slope $\gamma = -3.6$). 
    }
  \label{fig:S_dyn_k_jam_fss_above}
\end{figure}

\begin{figure}[t]
 \epsfxsize=8.6cm
 \epsffile{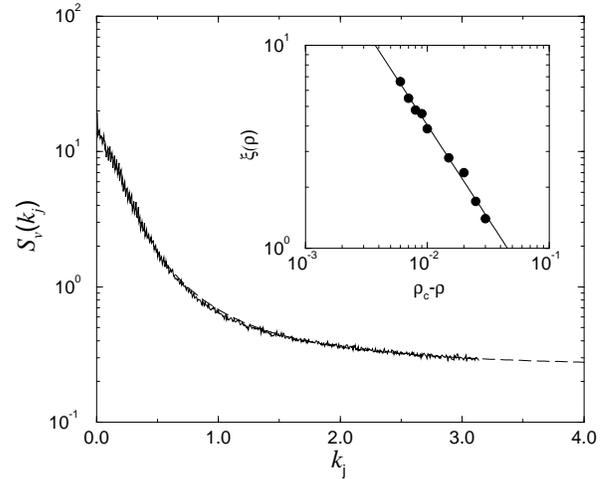}
  \caption{
    The dynamical structure factor $S_v(k, \omega )$ for
    $\omega/k = v_{\rm j}$, $v_{\rm max}=5$, $P=0.5$ and $N=1024$.
    The dashed line corresponds to a Lorentzian fit according to
    Eq.~(\ref{eq:lorentz_decay}).
    The inset shows the correlation length $\xi$ as a function of 
    $\rho_{\rm c}-\rho$.
    The obtained power-law behavior (solid line) agrees with 
    Eq.~(\ref{eq:nu}). 
    }
  \label{fig:xi_n_rho}
\end{figure}

In the following we carry out a quantitative analysis of the ridges to
obtain information about correlations between jammed particles.
We will show that among those particles long range correlations 
exist only above the transition whereas below the transition these
correlations are restricted to a finite correlation length.  
In Fig.~\ref{fig:S_dyn_k_jam_fss_above} we plot the 
values of the dynamical structure factor
$S_v(k, \omega)$ for $\omega/k=v_{\rm j}$, i.e.,~the 
values along the ridge which correspond to the 
jam modes.
In order to take finite-size effects
into account, we use the scaling ansatz
\begin{equation}
\left . S_v(k, \omega ) \right |_{\omega / k= v_{\rm j}}  \; =
\; N^{- \theta} \, f(N^{\zeta} k) .
\label{eq:S_k_jam_fss}
\end{equation}
For $\theta =-1.22$ and $\zeta=0.34$ we obtain a convincing
data collapse of all curves. 
The dynamical structure factor decays algebraically,
\begin{equation}
\left . S_v(k, \omega ) \right |_{\omega / k= v_{\rm j}} 
\; \sim \; k^{-\gamma},
\label{eq:allgebraic_decay}
\end{equation}
where the exponent is given by $\gamma\approx 3.6 \pm 0.2$.
This algebraic decay of the dynamical structure
factor indicates that the corresponding
correlation function is also characterized by an 
algebraic decay, i.e.,~the system displays 
long range correlations above the transition.
Next we are interested in the values of $S_v(k, \omega)$ for
$\omega/k = v_{\rm j}$ below the critical density.
These values are shown in Fig.~\ref{fig:xi_n_rho}.
Due to the finite amount of standing cars below the transition, the
jam mode does not vanish.
Our analysis shows that the jam mode decreases like a Lorentz
curve,  
\begin{equation}
  \left . S_v(k, \omega ) \right |_{\omega / k= v_{\rm j}} 
  \; \sim \; \frac{1}{1+(k\,\xi)^2 } + c \, .
  \label{eq:lorentz_decay}
\end{equation}
Herein $\xi$ denotes a correlation length, defined in the
velocity-particle space. 
The term $c$ takes into consideration that, caused by the
free flow phase, the jam mode does
not tend to zero for large $k$ but to a finite value.  
Beside the jam mode, Fig.~\ref{fig:xi_n_rho} shows a Lorentz
curve according to Eq.~(\ref{eq:lorentz_decay}), which has been fitted
to the data. 
To estimate the influence of finite-size effects on the
correlation length, we first determined the $\xi$ for different
values of $N$ (with $T=N$) and found that $\xi$ is almost in-affected
by finite-size effects for the densities shown in
Fig.~\ref{fig:xi_n_rho}.
Therefore, these values of the correlation length equal within
negligible errors those values of the correlation length found in the
limit $N\to\infty$ (with $N=T$). 
The dependence of the correlation length on $\rho_{\rm c}-\rho$ is
shown in the inset of Fig.~\ref{fig:xi_n_rho}. 
With $\rho$ approaching $\rho_{\rm c}$ the correlation length
diverges as 
\begin{equation}
  \xi \;  \sim \; (\rho_{\rm c}-\rho)^{-\nu} ,
  \label{eq:nu}
\end{equation}
with $\nu=0.92\pm0.05$.

Since $\xi$ describes the correlation of particles of the jammed
phase, the occurrence of long range order in the system at
$\rho=\rho_{\rm c}$ indicates that the system
displays criticality.  
As mentioned above, the correlation length $\xi$ is infinite above the
transition (long range order) and finite below. 
This coincides with measurements of the life time distribution of jams
at the critical point~\cite{NAGEL_2} which displays a power-law
behavior, i.e.,~traffic jams occur on all time scales. 
Below the critical density the life time of the jams is finite and no
long range correlations of jams can occur.

\begin{figure}[b]
 \epsfxsize=8.6cm
 \epsffile{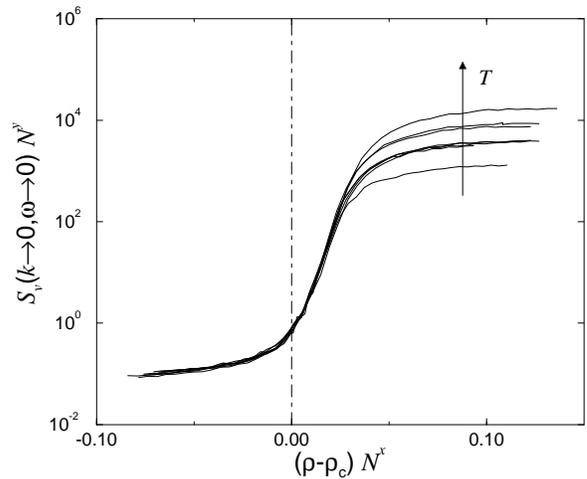}
  \caption{
    The smallest mode of $S_v(k\to 0, \omega\to 0)$, with
    $\omega/k=v_{\rm j}$, for different values of $N$
    ($N\in \{128,256,512,1024\}$) and  ratios $N/T$
    ($N/T\in \{0.5, 1, 2, 4\}$) and $P=0.321$. 
    \label{fig:S_v_k_mode_fss}} 
\end{figure}

In the following we address the question how long range order occurs
by approaching the transition point. 
Therefore we consider the smallest positive
mode on the ridge of jammed particles,
$S_v(k\to 0, \omega \to 0)$ with $\omega/k=v_{\rm j}$.
This value contains the information how long range and long time
correlations appear when the transition takes place and
it is believed to be 
closely related to an order parameter
(see for instance~\cite{SCHMITT_1}).
If the transition from the free flow regime
to the jammed regime can be described as a
critical phenomenon, $S_v(k\to 0, \omega \to 0)$
should vanish below the transition ($\rho<\rho_c$).
Simulating finite system sizes this means that the
smallest jam mode obeys the finite-size
scaling ansatz
\begin{equation}
  S_v(k\to 0, \omega \to 0) 
  = 
  (NT)^{-y/2} \, f[(NT)^{x/2} (\rho-\rho_{\rm c})],
  \label{eq:S_v_to_zero_fss}
\end{equation}
with $\omega / k= v_{\rm j}$.
We measured $S_v(k\to 0, \omega \to 0)$ for various
values of $N$ and $T$ and for $P=0.321$
which corresponds to a jam 
velocity $v_j\approx-1/2$~\cite{LUEB_7A}. 
The finite-size scaling works and for $y \approx -0.2$ and
$x \approx 0.1$ we got a good data collapse, which is plotted in 
Fig.~\ref{fig:S_v_k_mode_fss}.
Especially we obtain the data collapse for different ratios of $N$ and
$T$,
which shows that the system behaves isotropic in $N$ and $T$.
To investigate the dependence of the exponents on $P$,
we determined them also for another value of the noise parameter
($P=0.519$ corresponding to $v_j\approx-1/3$~). 
Due to the size of the corresponding error bars no significant $P$
dependence of the exponents could be observed.
Further investigation with improved accuracy are needed to clarify
this point.

Note that it is not justified to identify the
scaling exponents [Eq.~(\ref{eq:S_v_to_zero_fss})] with the usual
critical exponents of second order phase transitions ($y=\beta/\nu$
and  $x=1/\nu$).
First it is not clear if  
$S_v(k\to 0, \omega\to 0)$
equals the order parameter (see~\cite{SCHMITT_1} and
references therein). 
The second point is that the usual finite-size scaling ansatz rests on
the validity of the hyperscaling relation between the exponents (see
for instance~\cite{BINDER_1}). 
Despite these restrictions the finite-size scaling analysis of the
smallest jam mode reveals that it vanishes below the transition.
In the hydrodynamic limit ($N\to \infty, T\to \infty$)
no long range correlations in space and time 
occur for $\rho<\rho_c$.
Above the critical value the correlation function displays an
algebraic decay.   
Since the correlations of the jams are finite below and infinite above
$\rho_c$\,, the system displays critical behavior.
This agrees with the above mentioned investigations of the life time 
distribution~\cite{NAGEL_2} of jams and with measurements of the
relaxation time of the system. 
It was shown that the relaxation time~$\tau$ 
diverges at the critical
point with increasing system size $\tau \sim L^z$
with a $P$ depending exponent~$z$~\cite{CSANYI_1,SASVARI_1,EISEN_2}.
But one has to mention that
above the transition the measurements of the 
relaxation time yields unphysical result,
in the sense that the relaxation time becomes
negative~\cite{EISEN_2}.
We think that the origin of this behavior is 
caused by the inhomogeneous character of the
system above the transition where the system
separates into two coexisting phases.
The relaxation time measurement does not
take this inhomogeneous character into account.

\section{Conclusions}
\label{sec:conl}

We studied numerically the Nagel-Schreckenberg traffic flow model. 
The investigation of the dynamical structure factor allowed us to examine 
the transition of the system from a free flow regime to a jammed regime. 
Above the transition the dynamical structure factor exhibits two 
modes corresponding to the coexisting free flow and jammed phase. 
Due to the sign of their characteristic velocities $v_{\rm f}$ and 
$v_{\rm j}$,
both phases can clearly be distinguished. 

The analysis of the dynamical structure factor of the
velocity-particle space shows that above the transition the system
exhibits long range correlations of the jammed particles and below
the transition the correlations display a finite correlation length
that diverges at the critical point, indicating that a continuous 
phase transition takes place. 
Using a finite-size scaling analysis 
we showed that in the hydrodynamic limit 
the smallest jam mode which corresponds 
to the long range correlations of jams vanishes
below the transition.
We think that an extended investigation of this
quantity could lead to a convincing definition
of an order parameter which describes
the transition from the free flow to the jammed
regime.

\end{document}